\documentclass{emulateapj}
\usepackage{graphicx}
\usepackage{bm}
\usepackage{epsfig}
\usepackage{multirow}
\usepackage{color}
\usepackage{subfigure}
\usepackage{ulem}

\def\hideprivate{\global\long\def\private##1{\iffalse ##1 \fi}}
\hideprivate

\shorttitle{Early Years of EM-GW searches}
\shortauthors{Kasliwal \& Nissanke}

\begin{document}

\title{On Discovering Electromagnetic Emission from Neutron Star Mergers: \\
The Early Years of Two Gravitational Wave Detectors}
%Promising Prospects of Previously Perilous

\author{Mansi M. Kasliwal\altaffilmark{1} \& Samaya Nissanke\altaffilmark{2}}

\altaffiltext{1}{The Observatories, Carnegie Institution for Science,
  813 Santa Barbara St Pasadena, California 91101}

\altaffiltext{2}{Theoretical Astrophysics, California Institute of
  Technology, Pasadena, CA 91125, USA}

\begin{abstract}
We present the first simulation addressing the prospects of finding an electromagnetic (EM)
counterpart to gravitational wave detections (GW) during the early
years of only two advanced interferometers. The perils of such a search 
may have appeared insurmountable when considering the coarse ring-shaped GW localizations 
spanning thousands of deg$^{2}$ using time-of-arrival information alone. 
We show that leveraging the amplitude and phase information of the predicted GW signal 
narrows the localization to arcs with a median area of only $\approx$\,250 deg$^{2}$, thereby making an EM search tractable.  Based on the
locations and orientations of the two LIGO detectors, we find that the GW sensitivity is limited to one
polarization and thus to only two sky quadrants. Thus, the rates of GW events with two interferometers
is only $\approx$40\% of the rate with three interferometers of similar sensitivity.  Another important implication of the sky quadrant bias is that 
EM observatories in North America and Southern Africa would be able to systematically respond to GW triggers several hours sooner 
than Russia and Chile. Given the larger sky areas and the relative proximity of detected mergers, 
1m-class telescopes with very wide-field cameras are
well positioned for the challenge of finding an EM counterpart. 
Identification of the EM counterpart amidst the even larger numbers of false
positives further underscores the importance of building a comprehensive catalog of foreground stellar sources, 
background AGN and potential host galaxies in the local universe.
\end{abstract}

%%%%%%%%%%%%%%%%%%%%%%%%%%%%%%%%%%%%%%%%%%%%%%
\section{Introduction}
\label{sec:intro}

%%%%%%%%%%%%%%%%%%%%%%%%%%%%%%%%%%%%%%%%%%%%%%

The advent of advanced ground-based interferometers this decade is expected to usher in the 
era of {\it routine} gravitational wave (GW) detection \citep{Barish:1999,Sigg:2008, Accadia:2011,Somiya:2012}.
Binary neutron star (NS) mergers are anticipated to be amongst the most numerous and strongest GW sources \citep{abadieetal:2010}.  
NS mergers are predicted to produce neutron-rich outflows and emit electromagnetic (EM) radiation 
across many wavelengths and timescales as the ejected debris
interacts with its environment --- gamma (e.g.,~\citealt{Eichler:1989,Paczynski:1991,Narayan:1992}),  optical (e.g.,~\citealt{Li:1998,Kulkarni:2005,Metzger:2010,Roberts:2011,Piran:2012,Rosswog:2013}), infrared (e.g., \citealt{Barnes:2013,Kasen:2013,Tanaka:2013,Grossman:2013}) and radio (e.g., \citealt{Hansen:2001,Pshirkov:2010,Nakar:2011}). 

The discovery and characterization of the EM counterparts to GW detections 
promises to unravel astrophysics in the strong field gravity regime. % and extreme curvature space-time.  
Moreover, such EM-GW events will serve as the litmus test for whether 
NS mergers are indeed the sites of r-process nucleosynthesis (and hence, responsible for 
producing half the elements heavier than iron including gold, platinum and uranium; e.g. \citealt{ls76,mc90}). %Literally,
The accompanying surge of excitement in preparation for this endeavor has been described as analogous to the ``gold rush"  \citep{Kasliwal:2013}. 

In \citealt{Nissanke:2013} (hereafter, Paper I), we undertook
  an extensive end-to-end simulation on how to identify
  the elusive EM counterpart of a GW detection of NS mergers. We started
  with simulated astrophysical populations of NS mergers, evaluated GW detectability 
and considered three critical steps: (1) GW sky
localization and distance measures using different worldwide networks
of three to five GW interferometers, (2) subsequent EM detectability by a slew of
multiwavelength 
telescopes, and (3) identification of the merger counterpart amongst a possible fog of astrophysical
false-positive signatures. 
We showed how constructing GW
volumes and local Universe galaxy catalogs, can help
identify and reduce the number of false-positives, thereby enabling a secure
EM identification. 

Paper I simulated mergers detected by a network of three to five GW interferometers. However, 
given projected timescales for construction of advanced GW interferometers, it appears that the early years 
(and possibly the first detections) could be limited to a network of only two LIGO interferometers \citep{aasi:2013}.

In this letter, we consider new observational challenges specific to 
a network of only two GW interferometers.  We derive GW localization arcs (\S2,\S3), 
simulate detectability of EM counterparts (\S4), discuss false positives (\S5) and
conclude with strategies for timely EM-GW identification of NS mergers (\S6).

%******************************************************
%******************************************************

\section{GW Method: Detection and Source Characterization}

%******************************************************
%******************************************************

As detailed in \S2 of Paper I, we construct an astrophysically-motivated population of $4 \times 10^4$ NS-NS binaries out to a limiting redshift $z =
0.5$. %, where each NS has a physical mass of $1.4\,M_{\odot}$. 
Parameters include: binary masses, luminosity distance $D_L$, inclination angle to the observer's
line-of-sight $\cos \iota$, GW polarization angle $\psi$, and sky
position ${\bf n}$ (where $\hat{{\bf n}} \equiv (\theta,\phi) $ is the unit vector pointing to a binary on the
sky from a fixed Earth coordinate system, $\theta$ is the colatitude
and $\phi$ is the longitude). We associate each binary
with a random orientation and sky position, and distribute
the mergers assuming a constant comoving volume density for $D_L > 200$
Mpc ($\Lambda$CDM, \citealt{komatsu09}) or using a B-band luminosity galaxy catalog (CLU; \citealt{KasliwalPhD:2011}) 
for $D_L < 200$ Mpc. 

Next, we select the NS mergers that are detectable with only the two LIGO
interferometers at positions $\bf{x}_H$ and $\bf{x}_L$ (the subscripts
denote the Hanford and Livingston sites, hereafter LIGO-H and LIGO-L).  GW detection
and source characterization methods use optimum matched filtering 
between GW predictions and simulated detector streams (see \S3 of
Paper I for details). The measured GW
strain $h_{M} $ at a particular detector $\bf{x}_H$ or $\bf{x}_L$ is
the sum of the two GW polarizations, $h_+$ and $h_{\times}$, each
weighted by their antenna response functions $F_{+,[\mathbf{H/L}]}$ and $F_{\times,[\mathbf{H/L}]}$, and
multiplied by a time-of-flight correction. The time delay of the signal between the detector and the
coordinate origin is given by $\tau_{[\mathbf{H/L}]} \sim \hat{{\bf n}} \cdot
  \mathbf{x}_{\mathbf{[H/L]}} / c$, where $c$ is the speed of
  light. $h_+$ and $h_{\times}$ are functions of $D_L$, $\cos
\iota$, masses, and the GW frequency $f$. 
The antenna responses, $F_{+,[\mathbf{H/L}]}$ and
$F_{\times,[\mathbf{H/L}]}$, depend on $\hat{{\bf n}}$ and $\psi$.  
%The antenna responses, $F_{+,[\mathbf{H/L}]} (\hat{{\bf n}},\psi)$ and
%$F_{\times,[\mathbf{H/L}]} (\hat{{\bf n}},\psi)$ depend on
%$\hat{{\bf n}}$ and $\psi$.  
Based on triangulation with three or more
  interferometers, the time delay factor dominates over amplitude
  effects in the GW waveform when
  reconstructing sky location errors for the majority of sources
  \citep{Nissanke:2011,Veitch:2012}.  
  
  For LIGO-H and LIGO-L, we assume two anticipated noise curves at mid and full sensitivity (the upper red and black lines in Figure~1 of \citealt{aasi:2013})
  and idealized noise.  
 %We assume the interferometers have Gaussian, independent noise and a low
%frequency cut-off of 10 Hz. %Critically here, we note that the LIGO
%Hanford and Livingston detectors have orientations constructed such
%they have identical GW polarization sensitivity. 
We define a binary to be GW detectable if its expected signal-to-noise ratio (SNR) at
each detector is $>$ 6.5 and its expected network SNR (the root-sum-square of the
individual SNRs) $>$ 12. Consistent with \S2.4 of Paper I, the term
{\it Net2a} denotes a LIGO-H and LIGO-L network using such
a coincident trigger, whereas {\it Net2b} corresponds to an expected network SNR
trigger of $>$ 8.5. 

To infer the binary's sky position, %luminosity distance and inclination angle, 
we explicitly map out the
  full nine dimensional posterior probability density function (PDF)
  using MCMC methods (see \S3 of Paper I and
    \citealt{Nissanke:2010}) and derive 2-D PDFs in ($\cos \theta$, $\phi$). 
    We took particular care to start
  each MCMC chain at random all-sky positions. 
  %We then project a subset of 2D PDFs in ($\cos \theta$, $\phi$) by marginalizing over all other parameters. 
  %\samaya{CUT: Details of the MCMC machinery used are given in \S3 of Paper I, \cite{Nissanke:2010}, and \cite{lewis02}.}

Finally, to better understand our MCMC derived measures, we also implement two
toy models using amplitude-only GW waveforms. The
first model incorporates only time-of-arrival information, whereas the
second incorporates a combination of
time-of-arrival and the detector antenna responses. Our second toy model assumes a 5-D GW waveform of the
form:  $h_{{\it T+F}}\sim  \exp^{i 2 \pi \tau f} \left[ F_{+} (\hat{{\bf
    n}},\psi) \frac{(1 + \cos^2 \iota)}{D_L}  \, + \, F_{\times} (\hat{{\bf
    n}},\psi) \frac{- 2 \cos \iota}{D_L} \right]$, where we take $ f = 100 \mathrm{Hz}$. By %assuming white Gaussian noise and 
simulating hundreds of noise realizations, we map out the likelihood
function for $(\cos \theta, \phi)$ for randomly orientated and
located binaries on the sky at different
SNRs$_{[\mathbf{H/L}]}$. 

%******************************************************
%******************************************************

\bigskip
\bigskip
\section{GW Results: distance, localization arcs, and sky sensitivity}

%******************************************************
%******************************************************

\begin{figure*}
\centering 
\subfigure[NS-NS mergers: Luminosity Distance]{\label{fig:dist}\includegraphics[width=0.8\textwidth]{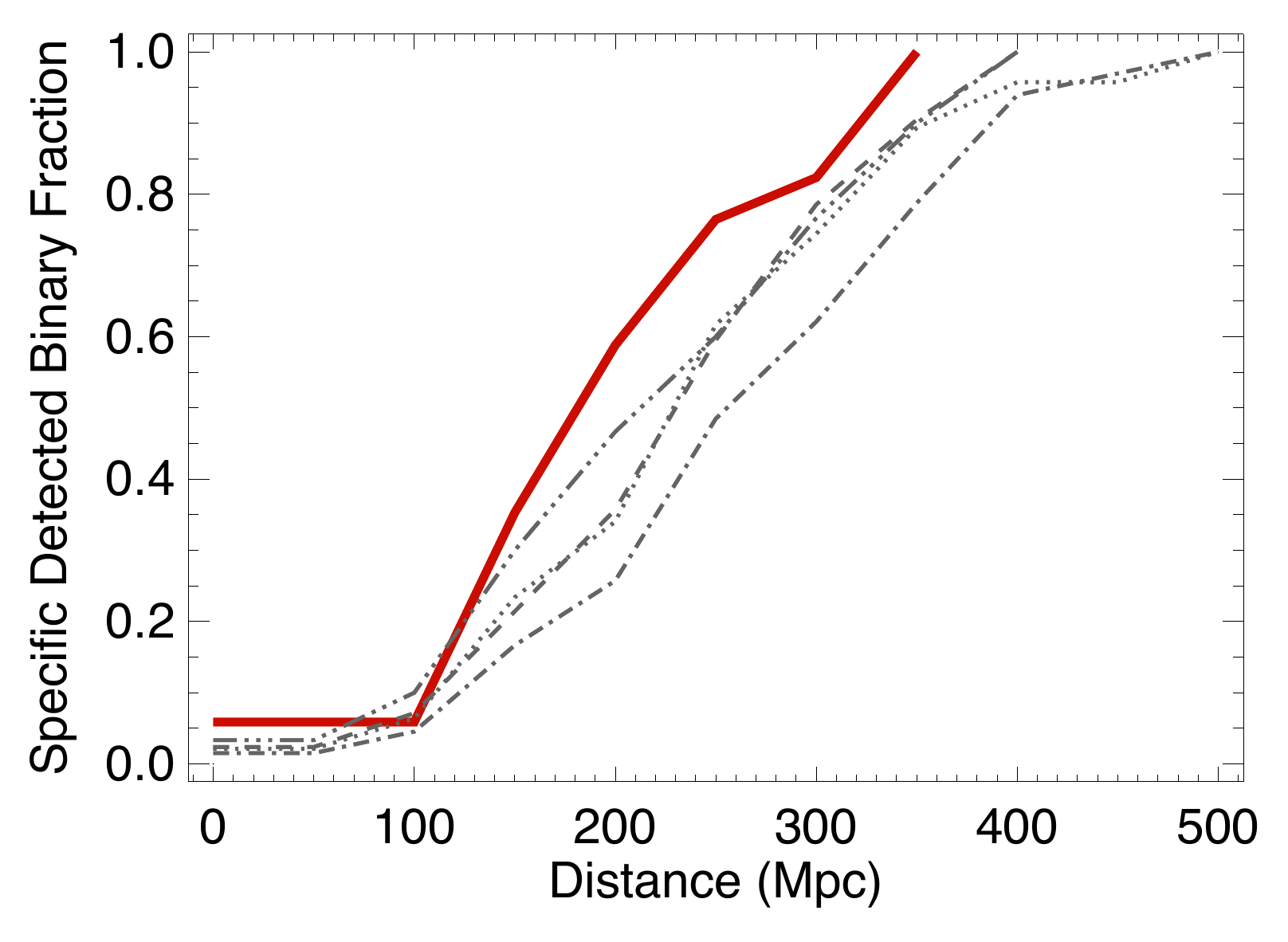}}
\vspace{0.1in}
\subfigure[NS-NS mergers: Sky errors]{\label{fig:area}\includegraphics[width=0.8\textwidth]{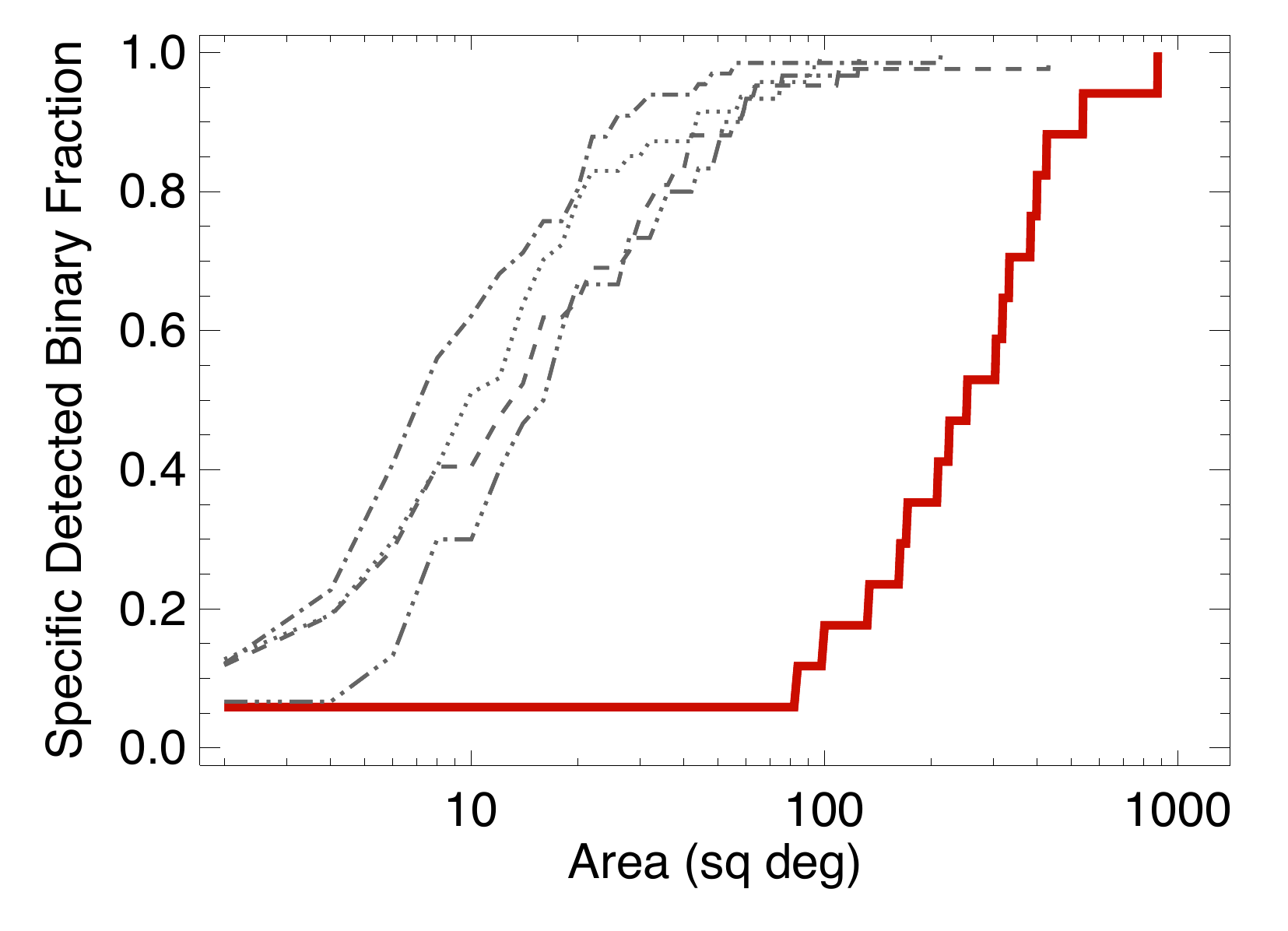}}
\caption{Cumulative distribution in luminosity distance
  (top panel) and 95\% confidence sky error (bottom panel) of NS-NS mergers. Red lines 
  denote a network of two GW interferometers. Gray lines denote Net 3--5 as presented in Paper I.   We require an 
  expected network SNR$>$12  and normalize to each specific network. %%Line-style
}
\label{fig:comparecum}
\end{figure*}

In Figure~\ref{fig:dist}, we show the cumulative distance distributions 
of NS mergers detectable using only LIGO-H and LIGO-L
at full-sensitivity. As expected, the distance distribution of mergers detected by Net2a 
is similar to those detected with Net3a-Net5a in \S2 of Paper I. At mid-sensitivity,
the distance distribution is scaled down by a factor of $\sim$ 0.6.   

In Figure~\ref{fig:area}, we show the cumulative histogram of sky
localizations at 95\% confidence regions (c.r.) for Net2a and
compare the distribution with that estimated by
Net3a--Net5a (\S2 of Paper I).  The median localization is 250
deg$^{2}$ compared with 17 deg$^{2}$ in Net3a. 
As in Paper I, we expect NS black-hole (BH) binaries to show a distribution similar to NS-NS.
At mid-sensitivity, we expect the specific distribution in sky
localizations to be similar to those at full-sensitivity 
because the majority of mergers will be detected at the SNR threshold
(distribution not shown here due to small number of detections).  
%since the geometric properties of inclination, sky position and the SNR distribution of the detected populations remains the same. 

\begin{figure*}
\centering 
\includegraphics[width=1.1\textwidth]{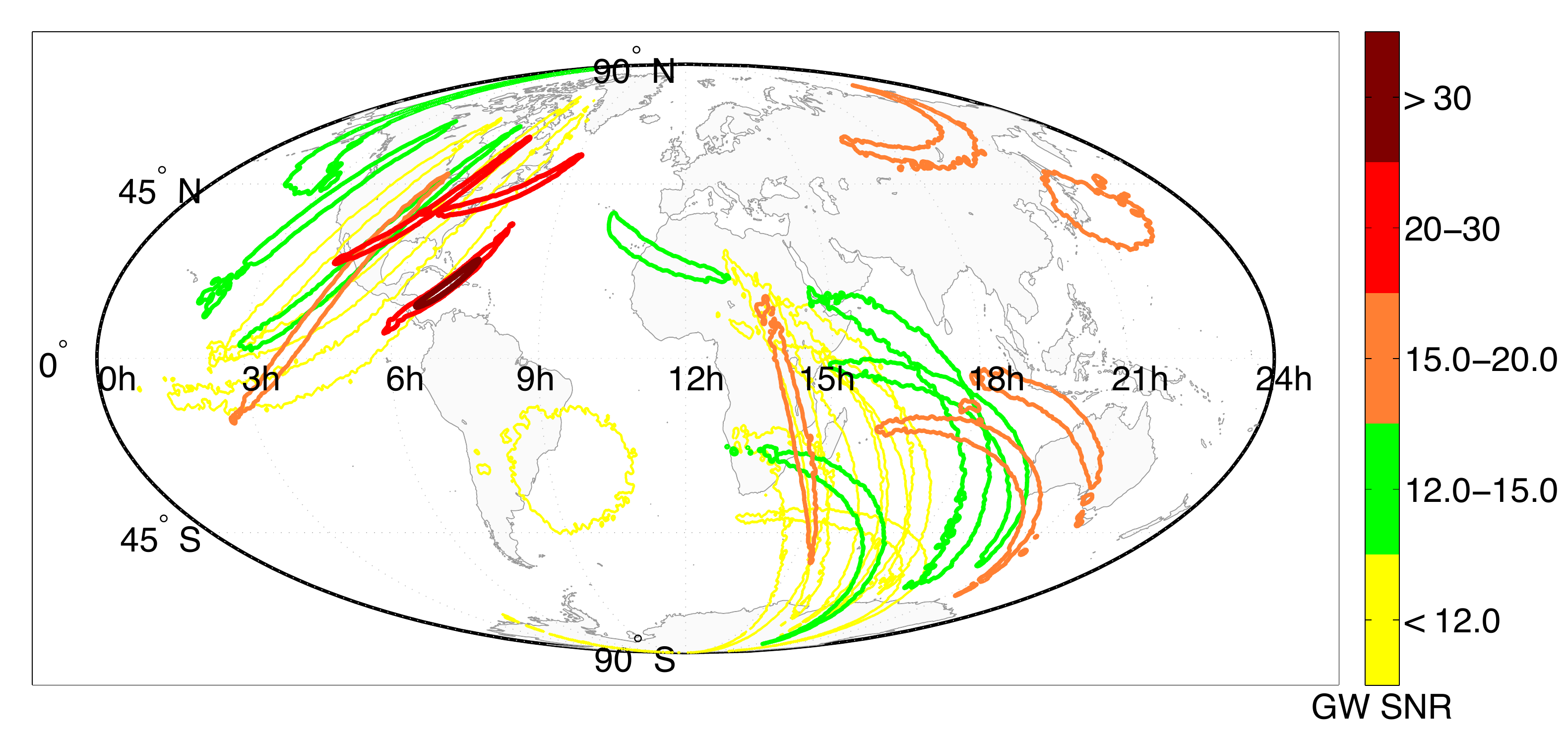}
\caption{Sky location and localization arcs of mergers detected by
  LIGO-H and LIGO-L. Color represents expected network SNR. 
Note that the quadrupolar antenna pattern has a bias towards two sky quadrants. The rate of detected mergers is 
$\approx$40\% of the rate of a three interferometer network.  The EM observatory location dictates a time lag in
response to GW trigger of up to to one day (Table~\ref{tab:em}).    
}
\label{fig:skymap}
\end{figure*}

In Figure~\ref{fig:skymap}, we show the
localization shapes, orientation and sky position of 
detected mergers at full-sensitivity. 
Using only time-of-arrival of signals at LIGO-H and LIGO-L,  
sky localization estimates have so far predicted annular error rings
for non-spinning mergers of several thousand deg$^{2}$
\citep{aasi:2013}. Instead, we
find that inclusion of $F_{+} (\hat{{\bf
    n}},\psi)$ and $F_{\times} (\hat{{\bf
    n}},\psi)$  in the %predicted
GW waveform's amplitude significantly improves localization errors to
arcs comprising several hundred deg$^{2}$. For Net 3--Net 5, we found that degeneracies between parameters 
%\sout{appearing only in the GW amplitude worsen the localization and even} 
result in non-contiguous areas for a handful of threshold mergers. %detectable sources 
\citep{Nissanke:2011}.  
Furthermore, we do not measure mirror-image arcs on the sky for any of the
detected binaries in our small sample. Indeed, for a single spinning NS-BH merger
using two initial LIGO sensitivities, \cite{Raymond:2009} generated a localization arc by including the BH's spin.

Investigations with our two toy models
improve our understanding of the MCMC results. % \sout{of arc-shaped error measures in isolated sky quadrants}. 
Using only time-of-arrival, averaging over a hundred noise realizations, we find GW localizations of almost annular rings
of 1000s deg$^{2}$. Adding detector antennae information to the same binary, we
find smaller GW arcs of $100$s deg$^2$  in only one sky quadrant as long as the network SNR $>$ SNR$_{\mathrm{crit}}$, 
where SNR$_{\mathrm{crit}}$ ranges from 8--12 
depending on orientation and sky position. Below SNR$_{\mathrm{crit}}$, the shapes depend on individual noise realizations
and we find mirror images of GW arcs in different sky
quadrants using $h_{{\it T+F}}$.

The quadrupolar antenna patterns of LIGO-H and LIGO-L
are oriented such that they are sensitive to identical GW
polarizations. Figure~\ref{fig:skymap} shows that
Net2 have significantly reduced sensitivity in two out of
four sky quadrants for sources arriving in the plane of the
interferometer arms. 
In contrast to Net3-5, we
do not find a strong correlation between the $D_L$ and sky error as a result of the two-quadrant
sky sensitivity. We find that two binaries at the same distance can have localization areas differing by an order of magnitude 
based on sky position.

Out of our underlying population,  %of 4$\times 10^{4}$ mergers, 
we find that only 17$\pm 4$ and 62$\pm 8$ mergers are detected in GWs using Net2a
and Net2b respectively. 
For the same population, we found that
43$\pm 7$ and 144$\pm 12$ mergers were detected using the corresponding Net3a
and Net3b respectively (Table~1 of Paper I). Therefore, Net2
will detect $\approx\,60 \%$ fewer mergers than Net3 using either SNR threshold. Using NS merger rate estimates %that span from 0.01 to 10 Mpc$^{-3}$ Myr$^{-1}$ 
 \citep{abadieetal:2010} and Eqn.~(7) in Paper I,
we expect 0.3--490 and 1.3--1640 mergers annually for Net2a and Net2b respectively. 

\section{EM Detectability (Triggered): \\ response-time, tiling and depth}
Our GW results, indicating a sky quadrant bias and coarse arc-shaped localizations, present new challenges for triggered 
EM follow-up. 
(The challenge for contemporaneous, independent detection in the $\gamma$-rays or X-rays
or low frequency radio is unchanged.) Given the median
localization of 250 deg$^{2}$ (at 95\% c.r.), we find the tiling is currently beyond the scope of existing infrared, ultraviolet and 
millimeter facilities. Hence, we consider follow-up by a representative set of optical facilities, with telescope apertures spanning 0.5--8\,m and 
camera angles spanning 2--50 deg$^{2}$  (see Table~\ref{tab:em}), and simulate relative detectability. 

%{\it Response-time:}  
Due to the Net2 sky quadrant bias (Figure~\ref{fig:skymap}),  mergers are preferentially detected 
overhead in the north and at hour angles around twelve in the south (relative to LST at LIGO-H/LIGO-L). 
Consequently, an EM observatory located around the same longitude as LIGO
can respond instantly if located in North America but only half a day later in Chile (see Table~\ref{tab:em}). This time-lag in response is 
critical for afterglows, which fade as a power-law in time, and some kilonova
models, which fade on few hours to day timescales. It is not relevant for radio facilities
looking for late-time emission on the months to year timescale. 

\begin{figure*}
\centering 
\includegraphics[width=0.8\textwidth]{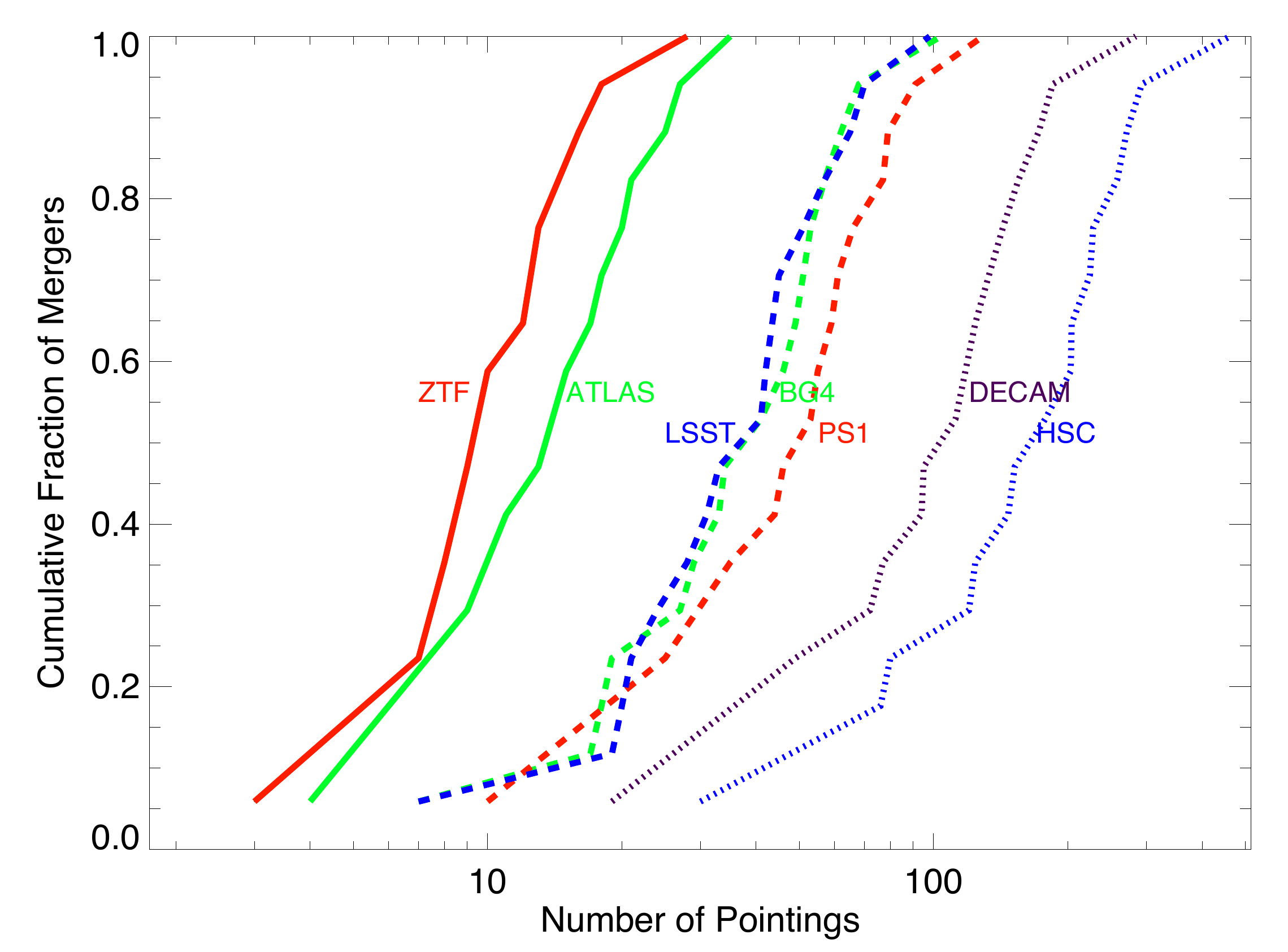}
\caption{Cumulative distribution of number of pointings necessary to
  tile localization arcs at all sky positions by LIGO-H and LIGO-L. Color represents telescope diameter: 0.5m-class (green), 1m-class (red), 4m-class (purple) and 8m-class (blue).
Line style represents camera angle: few tens of deg$^{2}$  (solid), several deg$^{2}$  (dashed) and few deg$^{2}$ (dotted).  
}
\label{fig:pointings}
\end{figure*}

%{\it Tiling:} 
Due to the elongated arc-shape and the coarser localization of hundreds of deg$^{2}$,  tiling 
presents a major challenge. % for EM facilities and elevates the importance of camera angle in the depth-cadence-area trade-off.
We compute an optimal tiling pattern to cover the GW localization
contour (95\% c.r.) for each merger at each EM facility (Figure~\ref{fig:pointings}). %%Small number stats geographical positions
While the widest cameras need $<$ 20 pointings, other facilities need hundreds of pointings. 
Naive division of the localization area by the camera field of view grossly underestimates the actual number of pointings required.
This tiling inefficiency factor %($\frac{N}{N_{\rm min}}$)  
has a median value of 1.6 for BG4/HSC, 1.8 for DECAM, 2.0 for LSST/PS1, 
2.3 for ATLAS and 2.6 for ZTF.  The localization arcs have a median width of
6.5$^{\circ}$ (in agreement with time-of-arrival estimates
  e.g., \citealt{Fairhurst:2010}). %and a minimum of 3.7 deg.  
Narrow-angle cameras can tile more nimbly than wide-angle cameras
(e.g., the BG4 tiling is 30\% more efficient than the contiguous PS1).
%Figure~\ref{fig:pointings} shows the total number of
%telescope pointings required as a cumulative histogram for each facility.  

%{\it Depth}:  
%To simulate the relative fraction of binaries detected by each EM facility, we compute the depth attained in a fixed duration.
With the number of pointings in-hand for each binary and for each EM facility, we compute the maximum exposure time (and hence, depth) allowable in a fixed duration. 
We assume three epochs (dithered to cover chip gaps) of one hour each with a detection above 5$\sigma$ in at least two epochs as minimum criterion for EM detection.
We take into account overhead between exposures which is dominated by readout for large mosaic cameras. Given the distance to each 
binary in our simulation, we convert the apparent magnitude depth to a luminosity. 
%Figure~\ref{fig:detectability} plots the relative detectable fraction of mergers as a  function of EM counterpart luminosity for various facilities. 

\begin{figure*}
\centering 
\includegraphics[width=\textwidth]{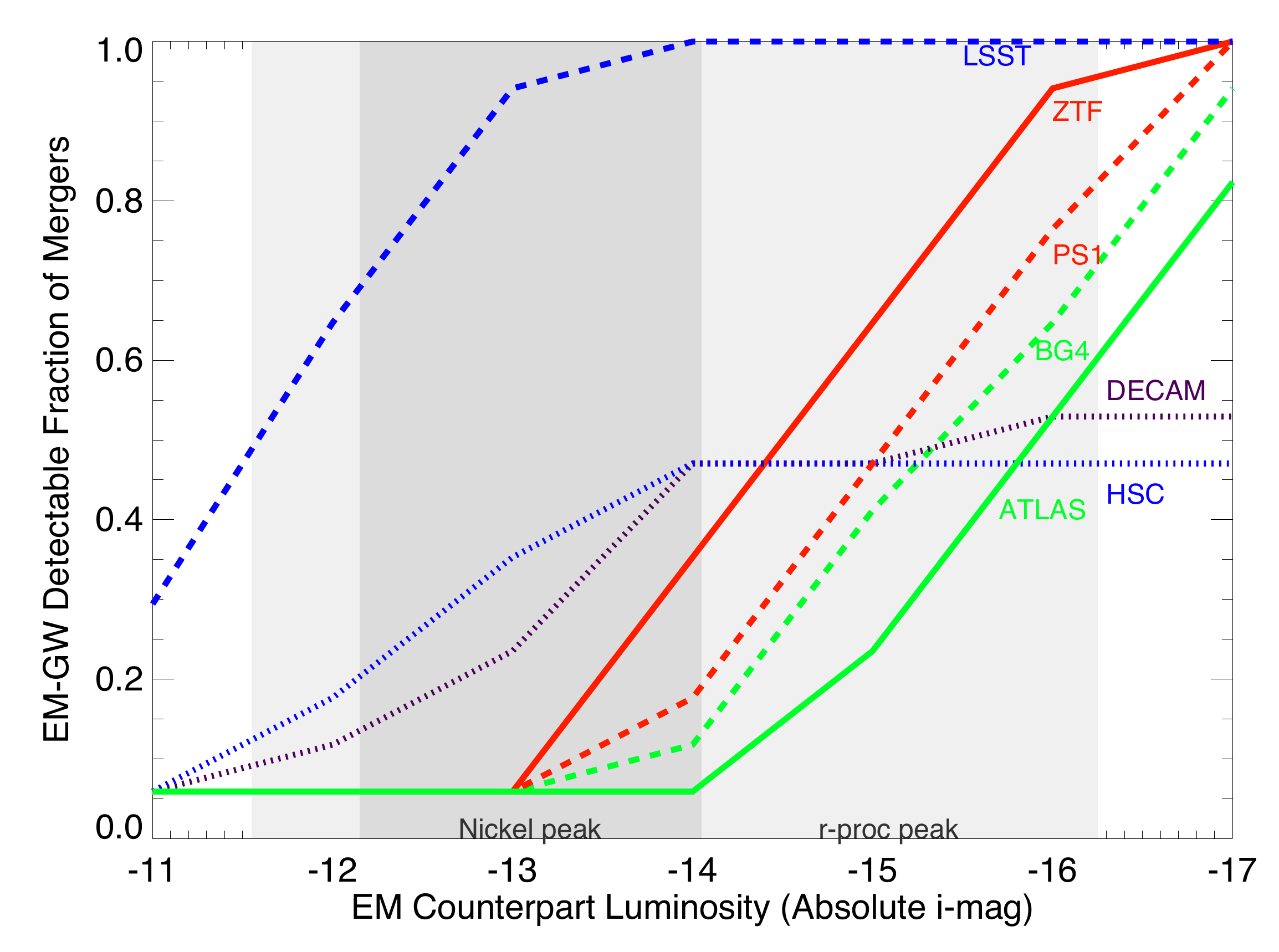}
\caption{Fraction of mergers detectable by a given EM facility as a function of kilonova luminosity (expressed in absolute i-band AB magnitude). Color and line-styles are same as in Figure~\ref{fig:pointings}. Shaded regions denote theoretical predictions for kilonovae \citep{Barnes:2013,Kasen:2013} ---  r-process powered peak (light grey;  M$_{\rm ejecta}\,\approx\,$ 10$^{-1}$--10$^{-3}$ M$_{\odot}$, v$_{\rm ejecta}\,\approx\,$ 0.1c--0.3c) and Nickel-56 powered peak (dark grey; M$_{\rm ejecta}\,\approx\,$ 10$^{-2}$--10$^{-3}$ M$_{\odot}$). 
All fractions are relative as an accessibility window of three hours with clear weather is assumed for all binaries at all facilities and no correction is made for lag in response (Table~1).  
}
\label{fig:detectability}
\end{figure*}

Our detectability simulation results for Net2 are very
different from those in Paper~I for Net3--5. Figure~\ref{fig:detectability} shows that small telescopes with large camera angles (e.g., ZTF) are {\it more} competitive 
than large telescopes with small camera angles (e.g., HSC)  for
detecting counterparts with an $i$-band luminosity brighter than $M_i=$
$-$14.5\,mag. Recent tantalizing near-infrared excess observed in one short GRB suggests bright kilonovae may be plausible (e.g., \citealt{Tanvir:2013}). %ejecta masses between 10$^{-2}$ to 10$^{-1}$ M$_{\odot}$
Note that response-time is not folded into this figure as there is a diversity in kilonova models ranging from some that
rise by 1\,mag and others that decline by 1\,mag in the first 12\,hours \citep{Barnes:2013}. 

%The wide camera angle of  ZTF is best suited to detecting kilonovae with luminosity brighter than $-$14.5\,mag. 
%Despite the smaller telescopes, ATLAS performs comparably to BG4 given its wider camera. The narrower angle of DECAM/HSC  leaves zero time for exposures 
%(given the product of overhead and number of pointings) for half the binaries. For the better localized binaries, even with a very short exposure of a few seconds, 
%DECAM/HSC are sensitive to low luminosity kilonovae.  
  
Finally, %we remind the reader that 
EM follow-up is further hampered by weather, sunshine, lunation and visibility window. %instrument availability. 
Thus, as discussed in \S5.3 of Paper I, we emphasize that all detectable 
fractions presented here should be interpreted as relative. %and discounted by a factor of four.    
 
%Overall, we note that binaries that are closer are preferentially better localized. However, the sky quadrant of the binary does matter. For example, binary 114 in our simulation
%is at a distance of 88\,Mpc but only has an SNR of 10.5 and a poor localization of 841 deg$^{2}$.

\section{EM Identification: False Positives}

Optical detection of candidate EM counterparts in a single epoch is only the first step. Multiple epochs are essential to distill the true EM counterpart from thousands of
astrophysical false positives in the foreground (e.g., moving objects in solar
system, variable stars in Milky Way) and background (e.g., supernovae and AGN in higher redshift galaxies).  An ongoing survey of the same
sky location to a similar depth would provide a historic baseline of variability of unrelated sources and serve as a severe filter.
%solution to the false positive problem.

Timely identification of the EM counterpart is critical for obtaining spectroscopic and multi-wavelength follow-up before the transient fades.  
In \S6 of Paper I, we considered five illustrative case studies to
quantify the false positive challenge and solutions in various scenarios.  
Here, we revisit the same five binaries in the context of  Net2. 

{\it A Beamed Merger (391\,Mpc):}  On account of the sky quadrant bias, this merger is not detected
by Net2 despite being beamed towards us. Given the lower rate of mergers detected with Net2 and the small 
fraction that is beamed  ($<$2.5\% for opening-jet angles $<$ 12$^{\circ}$),
we may not have the luxury of  the relatively easier search for the EM counterpart of a beamed
merger in the early years of Net2.

{\it A Close-in Merger (69\,Mpc):} Net2 localizes this merger to 23 deg$^{2}$ at full sensitivity and 
32 deg$^{2}$ at mid-sensitivity. This is a factor of $\approx$ 40 -- 50 coarser than Net 3. 
Thus, the number of false positives would be proportionately larger and it is even more important to have a complete catalog of nearby galaxies.
The fraction of ``golden" binaries that
are closer than 100\,Mpc remains $\approx$ 10\%.

{\it A High Galactic Latitude Merger (139\,Mpc):}  While Net 3 localized this 
merger to 19.5 deg$^{2}$, Net2 localizes this to 223
deg$^{2}$ at full sensitivity (it is not detected at mid-sensitivity due to its distance). 
Therefore,  there are ten times more background sources and it is even more important to have a complete catalog of nearby
galaxies and AGN variability. 

{\it A Low Galactic Latitude Merger (125\,Mpc):}  While Net 3 localized this merger to 1.8 deg$^{2}$,
Net2 localizes this to 100 deg$^{2}$ and 810 deg$^{2}$ at full and mid-sensitivity respectively. Thus, the foreground is 
55 times larger and it is even more important to build a catalog of stellar sources. 

{\it A Galaxy Cluster Merger (115\,Mpc):}. On account of the sky quadrant location, this merger is not detected 
by Net2 despite being relatively nearby. % (at full or mid-sensitivity).

%83.6/240.6, 98.8/313.6, 23.0/24.6, 225.5/177.8
%Case study II -- binary 206 - 23 sq deg (cf 0.6 sq deg)  
%Case study III--binary 201 - 98.8 sq deg (cf 1.8 sq deg)
%Case study IV--binary 210 - 225.5 sq deg (cf 19.5 sq deg)
%Case studies 1 and 5 were not detected!

\newpage

\section{Discussion}

The EM-GW challenge for NS mergers is three-fold: the GW localizations are {\it wide} (few hundred deg$^{2}$) and 
the predicted EM counterparts are {\it faint} (M$_{\rm i}\approx\,-$12 to $-$16\,mag) and {\it fast} (few hours to few days).  
With LIGO-H and LIGO-L, we derive arc-shaped localizations with a median area of 250 deg$^{2}$ 
that are biased to only two sky quadrants. The rate of GW-detectable mergers 
is $\approx$40\% of the rate of Net3 and the median localization area is 15 times coarser.

Strategies to maximize the odds of identifying faint and fast EM emission in wide GW arcs include:

\begin{itemize}

\item[-] A network of small ($<$ 1m) telescopes, despite the shallow depth, can leverage observatory location 
and wide-field to maximize rapid response to find bright and fast-evolving EM emission. Given the GW sky quadrant bias, North America and 
Southern Africa are recommended as the best locations for rapid response to Net2 triggers. 

\item[-] A medium (1m--3m) telescope, despite the medium depth,  can leverage an extremely large camera angle of few tens of deg$^{2}$ 
to be best positioned for searching for EM counterparts brighter than M$_{\rm i} < -$14.5\,mag. A dedicated facility with an ongoing survey to develop a baseline
of historic variability is recommended. 

\item[-] A large ($>$ 4m)  telescope, even with a relatively narrow camera angle of few deg$^{2}$, is uniquely
positioned to find faint EM counterparts. A planned large time investment, facilitation of camera availability and minimization of overheads between pointings 
are recommended to be able to efficiently tile a larger fraction of mergers.    

\end{itemize}

Independent of telescope size, the efficiency of a robust, real-time transient detection pipeline is an essential factor in assessing detectability. 
High quality image subtraction requires a deep pre-explosion reference image of the same sky location, preferably taken with the same EM facility.
Reliable candidate vetting needs a veteran machine learning algorithm, preferably trained on a large set of previous transient detections by the same EM facility. 
Thus, two facilities with identical hardware but disparate software would have different EM-GW detection capabilities.

Ongoing surveys have already successfully demonstrated the capability to discover optical transients which overcome the challenges of wide/faint/fast, but
one at a  time.
  For example, the discovery of an afterglow in 71
deg$^{2}$ addresses the {\it wide} challenge~\citep{Singer:2013}, the discovery of multiple transients spanning kilonova luminosities
addresses the {\it faint} characteristic (review in \citealt{Kasliwal:PASA2012}) and the discovery of a
relativistic explosion decaying on an hour timescale addresses the {\it
  fast} evolution \citep{Cenko:2013}.  Future surveys should prepare to simultaneously address all three challenges.   

In summary, the early years of a small number of coarse GW localizations will be challenging but tractable for an EM search. 
The combination of camera angle, telescope aperture, observatory location and survey software for each EM facility will delineate 
a different range in EM emission timescale and luminosity. A multi-pronged EM search would provide robust constraints on the vast 
phase space of kilonovae (ejecta mass, velocity and composition).  The findings of early searches will help
plan EM-GW identifications to a larger number of better localized mergers in the era of three to five GW interferometers. 

\bigskip

We thank C.~Hirata and E.~S.~Phinney for careful reading of the manuscript. 
We acknowledge valuable discussions with E. Bellm, J. Bloom, Y. Chen, 
J. M. D\'{e}sert,  A. Georgieva, P. Groot,  C. Galley, S. Mohta and D. Reitze.
We thank D. Kasen for making kilonova models available. 
MMK acknowledges generous support from the Hubble Fellowship and Carnegie-Princeton Fellowship. 
SMN is supported by the David \& Lucile Packard Foundation.

\newpage

%%%%%%%%%%%%%%%%%%%%%%%%%%%%%%%%%%%%%%%%%%%%%%%
%Mansi EM optical telescope table

\begin{deluxetable}{lccccccc}
\tabletypesize{\tiny}
%\tablewidth{0pt}
%\tablewidth{16.0cm}
%\tablecaption{Telescopes}
\tablehead{
\colhead{Facility} & 
\colhead{Aperture} &
\colhead{Field-of-View} &
\colhead{Exposure} & 
\colhead{Overhead} &
\colhead{Sensitivity}  &
\colhead{Detectable Fraction} &
\colhead{Lag} \\ 
%\colhead{Ref.} \\
\hline 
&
\colhead{ (m)} &  
\colhead{ (deg$^2$)} &
\colhead{ (sec) }  &
\colhead{ (sec) } &
\colhead{(5$\sigma$, i-mag)} &
\colhead{($-$16; $-$14; $-$12\,mag)} &
\colhead{(hr)}
}
\startdata
Palomar: Zwicky Transient Facility (ZTF)\tablenotemark{a}    & 1.2 & 47  &  600 &15  & 22.2 & 0.94; 0.35; 0.06 &   1 $\pm$ 2\\
BlackGEM-4 (BG4)\tablenotemark{b}  & 4$\times$0.6 & 4$\times$2  &  600 & 15  & 22.2 &  0.65; 0.12; 0.06 & 12 $\pm$ 2 \\
Pan-STARRS1 (PS1)\tablenotemark{c}                            & 1.8 & 7.0 & 180  & 10 & 21.9 &  0.76; 0.18; 0.06 &  3 $\pm$ 2 \\
ATLAS\tablenotemark{d}&  0.5 & 30  &  600 & 5  & 21.0 & 0.53; 0.06; 0.06 & 3 $\pm$ 2 \\
CTIO: Dark Energy Camera (DECAM)\tablenotemark{e} & 4.0 & 3.0 & 10 & 30  & 22.8 & 0.53; 0.47; 0.12 & 12 $\pm$ 2   \\  
Subaru: HyperSuprimeCam (HSC)\tablenotemark{f} & 8.2 & 1.77 & 1 & 20 & 22.4 & 0.47; 0.47; 0.18  & 3 $\pm$ 2  \\
Large Synoptic Survey Telescope (LSST)\tablenotemark{g} & 8.4 & 9.6 & 1 & 2 & 22.4 & 1.00; 1.00; 0.65 & 12 $\pm$ 2
\enddata
\tablenotetext{a}{\citealt{k12}, E. Bellm priv. comm.}
\tablenotetext{b}{P. Groot priv. comm., BG plans up to 20 telescopes, see https://www.astro.ru.nl/wiki/research/blackgemarray}
\tablenotetext{c}{http://pan-starrs.ifa.hawaii.edu} 
\tablenotetext{d}{J. Tonry priv. comm.} %\tablenotetext{d}{http://www.fallingstar.com/technical.php}
\tablenotetext{e}{D. DePoy priv. comm., \citealt{bkk+12}}
\tablenotetext{f}{http://www.naoj.org/cgi-bin/img\_etc.cgi}
\tablenotetext{g}{\citealt{aaa+09}}
\label{tab:em}
\end{deluxetable}
\vspace{0.2in}

%\bibliography{EMGW2}

\end{document}